# Thermal theory of aluminum particle ignition in continuum, free-molecular, and transition heat transfer regimes

Alexandre Ermoline

# Thermal theory of aluminum particle ignition in continuum, free-molecular, and transition heat transfer regimes

Alexandre Ermoline[a)]
Otto York Department of Chemical, Biological and Pharmaceutical Engineering, New Jersey Institute of Technology, Newark, New Jersey 07102, USA

Most studies on nano- and micro- sized aluminum particle ignition have been focused on the processes occurring inside particles. In the current paper, thermal ignition of an aluminum particle in the air is simulated with different heat transfer models: continuum, free-molecular, and Fuchs model. A single parabolic oxidation law is assumed in the particle size range from nano- to millimeter diameters. A particle is considered ignited when it reaches the oxide melting point. The criterion defining the limits of validity for each model is the ratio of continuum and free-molecular heat transfer rates. The dependence of ignition temperature $T_i$ on the particle size is in qualitative agreement with the experimental trends: $T_i$ can have values in the range of 700–1500 K for nanoparticles due to the dominating contribution of a free-molecular heat transfer, and sharp growth of $T_i$ with the particle size in the range of 1–100 $\mu$m diameter is due to the transitional character of heat transfer. For small values of the accommodation coefficient, ignition may occur in the critical ignition mode with the thermal runaway. The results suggest the importance of non-continuous heat transfer and, in particular, energy accommodation in ignition of nano- and micro- sized particles. *Published by AIP Publishing.* https://doi.org/10.1063/1.5039691

## I. INTRODUCTION

Aluminum has been used as an admixture in solid propellants, fuels, and explosives. During the last two decades, with the advent and development of nanotechnology, a lot of interest has been drawn towards aluminum nanoparticles. It was found that compared to the micro-sized particles, nanoparticles have lower ignition temperatures[1] and cause higher burning rates when added into propellants[2] and thermites.[3] Numerous experimental data on aluminum ignition covering the range of particle sizes from several nanometers to millimeters are available in the literature. At the same time, while ignition of large particles is well understood, no satisfactory theory exists for ignition of small aluminum particles having diameters of the order of several micrometers and less.

The general trend of aluminum ignition temperature vs. particle size is well established.[1,4] Coarse particles (greater than 10–100 $\mu$m diameter) have ignition temperatures slightly lower than the aluminum oxide melting point ($\sim$2300 K) with the very weak dependence on the particle size. For smaller particles, ignition temperature sharply declines when the particle size decreases, reaching values close to the aluminum melting point (933 K) for nanoparticles. Values lower than the aluminum melting point were reported, as low as 900 K[5] or 820 K—the onset of fast oxidation obtained in thermogravimetric and differential thermal analysis measurements.[6]

Friedman and Maček[7–9] studied ignition of particles in the diameter range of 15–65 $\mu$m and found that ignition temperature is close to the melting point of oxide. This has led the authors to believe that the cause of aluminum ignition is the increase in the particle oxidation rate as a result of melting of oxide covering aluminum particles. In other ignition studies,[10–15] the direct sample temperature measurements and observations showed that aluminum temperature at the ignition moment is lower than the oxide melting point. Also, the disintegration of the aluminum oxide layer was observed before ignition. These findings suggest that aluminum may ignite as a result of loss of oxide protective properties before oxide melts. The thermal ignition theory with the parabolic oxidation law[16,17] and fixed temperature as an ignition criterion—either oxide melting point or slightly lower temperature—can be used in this case.

This theory fails to describe ignition of particles having sizes on the order of several micrometers and less, for which experiments show lower ignition temperatures.[18,19] To address this discrepancy, Gurevich *et al*.[20] proposed the model where the possible effect of the presence of two different, crystallized and amorphous, polymorphs in aluminum oxide was considered. Gurevich *et al*. suggested that ignition temperature dependence is the result of competition between crystallization and non-protective oxidation over the amorphous oxide area. Rozenband and Vaganova[21] showed that cracks in the oxide film can be formed as a result of thermo-mechanical stresses caused by different densities and linear expansion coefficients of metal and oxide. These factors could lead to lower ignition temperatures for smaller particles. Based on the results of thermogravimetric analysis, Trunov *et al*.[22] developed the quantitative model of aluminum powder oxidation. Interpreting different stages in oxidation observed in thermogravimetric analysis as growth and transformation of different alumina polymorphs, Trunov *et al*. found parameters characterizing these processes and applied them to the particle ignition. According to this

[a)]Electronic mail: alexandre.ermoline@njit.edu

model, because of different densities of polymorphs, discontinuities in oxide layers are produced, leading to the enhancement of oxidation and self-heating with the subsequent ignition.

Experimental observations showed that the onset of nanoparticle oxidation either coincides or is above the aluminum melting point.[23–25] Therefore, aluminum melting is believed to play an important role in nanoparticle ignition. Rai et al.[25] developed the phenomenological model of nanoparticle oxidation with diffusion of reacting species through the oxide layer as a limiting process: oxygen diffusion—below melting point and diffusion of both oxygen and aluminum—above melting point. Because of differences in densities of molten and solid aluminum, internal pressure gradients are developed that may lead to thinning and rupture of the oxide film and enhancing diffusion of species. Large internal pressures inside nanoparticles were predicted in molecular dynamic simulations by Campbell et al.[26] Levitas et al. proposed the aluminum melt dispersion mechanism of ignition.[27,28] When aluminum melts, high tensile stresses are developed because of the particle volume change, reaching up to 4 GPa, according to calculations.[28] This pressure ejects small aluminum clusters out of a particle with the subsequent fast oxidation or ignition of these clusters. According to Levitas et al.,[29] this mechanism may be responsible for ignition at high heating rates.

In all the models considered above, the ignition behavior was explained by kinetics and accompanying processes inside particles without taking into account possible limitations of heat transfer or considering the heat transfer in the continuum approximation. The approximation of gas as continuum is valid for small Knudsen numbers $Kn \ll 1$. When $Kn \gg 1$, heat transfer occurs in the free-molecular regime, i.e., energy exchange between gas and particles is the result of individual molecule–particle collisions. At some particle sizes between these two extreme cases, transition heat transfer takes place, which has an intermediate nature between continuum and free-molecular approaches. The approximate range of transition heat transfer is typically estimated as $Kn \sim \mathcal{O}(1)$. For elevated temperatures in the air at 1 atm, particles having sizes of several micrometers and less can be expected to ignite in a transition regime. The free-molecular character of heat transfer increases for nanoparticles. Thermal ignition theory that has been successfully used for large particles should be modified to include the free-molecular heat transfer effects.

In this article, the thermal ignition model of an aluminum particle in a hot air is considered at Knudsen numbers ranging from $Kn \ll 1$ (continuum approximation) to $Kn \gg 1$ (free-molecular approximation). To describe transition heat transfer, the two-layer Fuchs model is used.[30] An extensive literature exists with the applications of this model to different types of heat and mass diffusion problems, including combustion.[31,32] The equations of this model approach asymptotically those of free molecular ($Kn \to \infty$) and continuum ($Kn \to 0$) models, so, in principle, the Fuchs model is applicable at any value of Kn, while the free-molecular and continuum regimes can be considered as its limiting cases. But the Fuchs model is difficult to analyze. Instead, the alternative approach is used here: first, results of continuum and free-molecular models are considered, and then, the Fuchs model results are treated as the corrections in the transition heat transfer regime. The limits of application of each model are established by comparison of the results of different models.

It is a generally accepted fact that large particles are oxidizing according to the parabolic oxidation law. This oxidation law is assumed here for all particle sizes. Historically, as a result of continuum heat transfer model deficiency, various kinetic models were developed aiming to explain the ignition temperature decrease for small particles. They typically involve parabolic oxidation with some additional factors that become important at some stage, such as oxide cracking, or large pressures developed inside the particles. It is of interest to see how ignition temperature depends on the particle size with the parabolic oxidation and proper accounting for the free-molecular heat transfer effects. The goal of this study is revisiting the thermal ignition theory with the free-molecular and transition heat transfer models and parabolic oxidizing kinetics. In accord with this goal, no modification of this kinetics due to any possible processes inside the particle is considered.

The key parameter in a transition and free-molecular heat transfer—thermal accommodation coefficient—is very poorly known for temperatures and materials of interest in combustion. It is varied in the simulations in the range from 1 to 0.01. This study is focused on the dependence of ignition temperature on the particle size, with the accommodation coefficient and initial aluminum oxide thickness as varying parameters.

## II. MODEL AND EQUATIONS

### A. Model description

A spherical particle, consisting of an aluminum core surrounded by an aluminum oxide layer, is placed into a hot quiescent air. The limiting factor in oxidation is the diffusion of reactants through the oxide layer, i.e., oxidation is described by the parabolic law. The oxygen concentration at the particle surface is the same as in the air far from the particle. Aluminum in the particle core undergoes phase transformation, melting or freezing, at temperature $T_{m,Al}$. Diameters of the particle and reacted core are related through the shrinking-core model.[33] Temperature distribution inside the particle is neglected. Heat exchange with the surroundings occurs due to heat conduction and radiation, and the Stefan flow is not taken into account. The emissivity of a particle is assumed constant. The quasistatic approximation is used: particle temperature changes slowly enough, so that the heat flux in the environment adjusts infinitely fast to the particle temperature. Ignition temperature is defined as the minimum ambient temperature when the particle attains the melting point of aluminum oxide. No possible size effects on thermophysical or kinetic parameters of nanoparticles are taken into account.

In the free-molecular model, the Maxwellian velocity distribution is assumed, i.e., the non-equilibrium effects caused by thermal gradients are neglected.

### B. General equations

At temperatures other than aluminum melting point $T_{m,Al}$, the energy balance at the particle is

$$c_p m_p \frac{dT_p}{dt} = -\Delta H \frac{dm_{ox}}{dt} - \dot{q} - 4\pi r_p^2 \epsilon \sigma \left(T_p^4 - T_\infty^4\right),$$
$$T_p \neq T_{m,Al}, \quad (1)$$

where $T_p$ is the particle temperature, $c_p$ is the particle specific heat at constant pressure, $r_p$ is the particle radius, and $m_p$ and $m_{ox}$ are the masses of a particle and aluminum oxide, respectively. $\Delta H$ is the enthalpy of reaction per mass of oxide, $\epsilon$ is the particle emissivity, $\sigma$ is the Stefan-Boltzmann constant, and $T_\infty$ is the ambient temperature. $\dot{q}$ is the heat flow rate, which is specified below for each particular model: continuum $\dot{q}_c$ and free-molecular $\dot{q}_{fm}$.

The energy equation during aluminum melting or freezing can be written as

$$-L m_{Al} \frac{df_s}{dt} = -\Delta H \frac{dm_{ox}}{dt} - \dot{q} - 4\pi r_p^2 \epsilon \sigma \left(T_p^4 - T_\infty^4\right),$$
$$T_p = T_{m,Al}, \quad (2)$$

where $m_{Al}$ is the aluminum core mass, and $L = H_{Al,l} - H_{Al,s}$ is the latent heat of aluminum melting. $f_s$ is the mass fraction of solid aluminum in the particle core

$$f_s = \frac{m_{Al,s}}{m_{Al}}, \quad (3)$$

where $m_{Al,s}$ and $m_{Al}$ are the masses of solid aluminum and the core, respectively. $f_s = 1$ for temperatures less than the aluminum melting point, and $f_s = 0$ for higher temperatures. At $T_p = T_{m,Al}$, $f_s$ changes from 1 to 0 during aluminum melting and from 0 to 1 during solidification.

The aluminum oxidation rate is

$$\frac{dm_{ox}}{dt} = \frac{4\pi r_{Al} r_p}{h} k_0 X_{O_2}^n \exp\left(-\frac{E}{RT_p}\right), \quad (4)$$

where $r_{Al}$ is the radius of the aluminum core, $k_0$ is the preexponent, $E$ is the activation energy, $X_{O_2}$ is the oxygen mole fraction, $n$ is the order of reaction with respect to oxygen, $R$ is the universal gas constant, and $h$ is the oxide layer thickness: $h = r_p - r_{Al}$.

Initial conditions for Eqs. (1), (2), and (4) are

$$t = 0: \quad T_p = T_{p0}, \quad f_s = f_{s0}, \quad m_{ox} = m_{ox,0}. \quad (5)$$

The variables are related through the following algebraic equations:

- The enthalpy of aluminum $H_{Al}$ during phase transition includes enthalpies of the solid and liquid phases

$$H_{Al} = f_s H_{Al,s} + (1 - f_s) H_{Al,l}. \quad (6)$$

- The particle specific heat $c_p$ includes specific heat of aluminum $c_{Al}$ and aluminum oxide $c_{ox}$

$$c_p m_p = c_{Al} m_{Al} + c_{ox} m_{ox}. \quad (7)$$

- The aluminum mass is related to the oxide mass according to the stoichiometry

$$m_{Al} = m_{Al,0} - \frac{2M_{Al}}{M_{ox}}(m_{ox} - m_{ox,0}), \quad (8)$$

where $m_{Al,0}$ is the initial mass of an aluminum core, and $M$ is the molar mass of the corresponding component.

- The core density $\rho_{Al}$ is the average density of two phases, solid and liquid, having densities $\rho_{Al,s}$, $\rho_{Al,l}$, respectively,

$$\rho_{Al} = \left(\frac{f_s}{\rho_{Al,s}} + \frac{1-f_s}{\rho_{Al,l}}\right)^{-1}. \quad (9)$$

- The particle core mass is

$$m_{Al} = \frac{4}{3}\pi r_{Al}^3 \rho_{Al}. \quad (10)$$

- The particle and the core radii are related according to the shrinking core model

$$\frac{4}{3}\pi \rho_{ox}\left(r_p^3 - r_{Al}^3\right) = m_{ox}. \quad (11)$$

Ignition temperature $T_i$ is the ambient temperature $T_\infty$ at which the net heat balance at the particle surface is zero at the oxide melting point $T_{m,ox}$. With the notation for the reaction heat release term

$$\dot{q}_r = -\Delta H \frac{4\pi r_{Al} r_p}{h} k_0 X_{O_2}^n \exp\left(-\frac{E}{RT_p}\right), \quad (12)$$

this condition can be written as

$$\dot{q}_r(T_{m,ox}, T_i) - \dot{q}(T_{m,ox}, T_i) - 4\pi r_p^2 \epsilon \sigma \left(T_{m,ox}^4 - T_i^4\right) = 0. \quad (13)$$

This condition is equivalent to the definition of ignition temperature as the minimum gas temperature for which a particle gets heated to the oxide melting point.

### C. Continuum model equations

The enthalpy of combustion in the continuum model is

$$\Delta H = H_{ox}(T_p) - \frac{2M_{Al}}{M_{ox}} H_{Al}(T_p) - \frac{3M_{O_2}}{2M_{ox}} H_{O_2}(T_p), \quad (14)$$

where $H$ is the enthalpy per mass of the corresponding component.

In spherical coordinates with the origin at the particle center, the steady-state heat equation in gas is

$$r > r_p: \quad \frac{d}{dr}\left(r^2 k \frac{dT}{dr}\right) = 0, \quad (15)$$

where $k$ is the thermal conductivity of air. The dependence of thermal conductivity on temperature is taken into account by the 1/2 rule,[34] according to which the thermal conductivity is calculated at the effective temperature

$$T = \frac{1}{2}(T_p + T_\infty). \quad (16)$$

The solution of Eq. (15) with the boundary conditions

$$r = r_p: \quad T = T_p, \quad r \to \infty: \quad T = T_\infty \qquad (17)$$

yields the continuum heat flow rate from the particle $\dot{q}_c$

$$\dot{q}_c = 4\pi k r_p (T_p - T_\infty). \qquad (18)$$

### D. Free-molecular model equations

The difference in the reaction enthalpy compared to the continuum case is that oxygen reacting with the particle has the enthalpy at the surrounding gas temperature $T_\infty$

$$\Delta H = H_{ox}(T_p) - \frac{2M_{Al}}{M_{ox}} H_{Al}(T_p) - \frac{3M_{O_2}}{2M_{ox}} H_{O_2}(T_\infty). \qquad (19)$$

The heat flow rate $\dot{q}_{fm}$ from the particle to the surrounding gas is

$$\dot{q}_{fm} = \sqrt{2\pi R_g}\, r_p^2 p \alpha \frac{\gamma^* + 1}{\gamma^* - 1} \frac{T_p - T_\infty}{\sqrt{T_\infty}}, \qquad (20)$$

where $R_g$ is the gas constant, $p$ is the pressure, $\alpha$ is the thermal accommodation coefficient, and $\gamma^*$ is the mean specific heat ratio defined according to Filippov and Rosner[35] as

$$\frac{1}{\gamma^* - 1} = \frac{1}{T_p - T_\infty} \int_{T_\infty}^{T_p} \frac{1}{\gamma - 1} dT. \qquad (21)$$

### E. Fuchs model equations

The major idea of the Fuchs model is the division of the entire space around a particle into two regions: the one closely surrounding the particle (called the Knudsen layer or the Langmuir layer) across which the energy transport occurs by freely moving molecules and the outside space, where the medium is treated as continuum. In each region, the corresponding heat transfer model, free-molecular or continuum, is used. At the boundary interface between these regions, which is located at the distance about the mean free path apart from the particle surface, the heat flow rates from the two regions are matched. More details can be found elsewhere.[30,35,36]

With the notation $r_\delta$ for the radial position of the Knudsen layer interface and $T_\delta$ for the interface temperature, the free-molecular heat flow rate can be written as

$$\dot{q}_{fm} = \sqrt{2\pi R_g}\, r_p^2 p \alpha \frac{\gamma^* + 1}{\gamma^* - 1} \frac{T_p - T_\delta}{\sqrt{T_\delta}}, \qquad (22)$$

where the mean specific heat ratio $\gamma^*$ can be found from

$$\frac{1}{\gamma^* - 1} = \frac{1}{T_p - T_\delta} \int_{T_\delta}^{T_p} \frac{1}{\gamma - 1} dT. \qquad (23)$$

These equations are identical to Eqs. (20) and (21), with the only difference of using $T_\delta$ in place of $T_\infty$. Similarly, we can write the heat flow rate in the continuum region as in Eq. (18) substituting $r_\delta$ and $T_\delta$ in place of $r_p$ and $T_p$, respectively,

$$\dot{q}_c = 4\pi k r_\delta (T_\delta - T_\infty), \qquad (24)$$

with $k$ calculated at temperature $T = 1/2(T_\delta + T_\infty)$. The temperature at the Knudsen layer interface $T_\delta$ is defined by the equality of the heat flow rates

$$\dot{q}_c = \dot{q}_{fm}. \qquad (25)$$

Liu et al.[36] showed using the theoretical expression for the Knudsen layer thickness by Wright[37] that the Knudsen layer thickness can be taken as the mean free path $\lambda$

$$r_\delta = r_p + \lambda. \qquad (26)$$

The Maxwell mean free path is calculated with the temperature at the Knudsen layer interface $T_\delta$, according to Liu et al.[36]

$$\lambda = \frac{4k(T_\delta)}{(9\gamma(T_\delta) - 5)p} (\gamma(T_\delta) - 1) \left( \frac{\pi m_g T_\delta}{2k_B} \right)^{1/2}, \qquad (27)$$

where $m_g$ is the average mass of a gas (air) molecule, and $k_B$ is the Boltzmann constant. Equations (22)–(27) constitute the Fuchs model. The heat flow rate $\dot{q}_{fm}$ from Eq. (22) [or $\dot{q}_c$ from Eq. (24)] is used in Eqs. (1) and (2).

Oxygen reacting at the particle surface comes from the Knudsen layer interface having temperature $T_\delta$, and the enthalpy of reaction is

$$\Delta H = H_{ox}(T_p) - \frac{2M_{Al}}{M_{ox}} H_{Al}(T_p) - \frac{3M_{O_2}}{2M_{ox}} H_{O_2}(T_\delta). \qquad (28)$$

### F. Applicability of quasi-steady approximation

The applicability of quasi-steady approximation for continuum and transition models was demonstrated by Filippov and Rosner[35] for the case of heating (or cooling) a dense solid particle in the gas by conduction. The time of particle heating is much larger than time of thermal equilibration in the gas due to higher specific heat and density of a particle. The heat reaction rates considered here are of the same order of magnitude as heat transfer rates. Therefore, justification of the quasi-steady approach by Filippov and Rosner can be directly applied to the current case. The details of rate estimation are provided in Appendix. These conclusions are valid in the absence or small radiation. But when radiation becomes dominating, the justification of the quasi-steady approach loses significance due to negligible contribution of conductivity to the heat transfer process.

## III. PARAMETERS AND SOLUTION PROCEDURE

The equations are solved for a particle with the initial temperature of 298 K placed in the hot air at 1 bar. The thermal conductivity and the specific heat ratio of air are calculated with the polynomial expressions presented by Liu et al.[36] for the temperature range of 300–4000 K. Temperature-dependent data for the specific heat of Al and $Al_2O_3$ and all the values of enthalpy are taken from the NIST database.[38]

The choice of kinetic parameters was dictated by the condition: the data had to be obtained for the bulk aluminum or large particle sizes. Merzhanov et al.[10] studied ignition of aluminum wire of 30–50 $\mu$m diameter in pure oxygen in the temperature range of 1600–2000 K and atmospheric pressure. The parabolic oxidation law was found to be

$$\frac{dh}{dt} = \frac{1.9 \times 10^{-5}}{h} \exp\left(-\frac{17 \text{ kcal/mol}}{RT}\right) \text{cm s}^{-1}. \quad (29)$$

The experimental data in Eq. (29) were obtained for $X_{O_2} = 1$, and to extend them to other concentrations, the right hand side of Eq. (29) was multiplied by a factor $X_{O_2}^n$. Also, the parameters were determined in the assumption of a flat aluminum shape, i.e., $h/r_p \ll 1$. Writing the rate of oxide mass formation in this approximation and using the experimental value in Eq. (29), we have

$$\frac{dm_{ox}}{dt} = \rho_{ox} 4\pi r_p^2 \frac{dh}{dt}$$
$$= 4\pi r_p^2 \frac{1.9 \times 10^{-5} \text{ cm}^2 \text{ s}^{-1} \rho_{ox} X_{O_2}^n}{h} \exp(-E/RT_p). \quad (30)$$

Comparing this expression with Eq. (4) (with $r_{Al} \to r_p$), we have $k_0 = 1.9 \times 10^{-5} \rho_{ox}$ g cm$^{-1}$ s$^{-1}$. The mole fraction of oxygen in the air $X_{O_2} = 0.21$. The first order of reaction is assumed: $n = 1$. It should be noted that some other values of preexponent and activation energy could be found in the literature. Since the accommodation coefficient is not known, there is no reason or clear criterion to prefer one kinetics over another. The numerical values of parameters used in calculations are listed in Table I.

The set of equations for continuum and free-molecular models consists of two differential equations, Eq. (1) [or Eq. (2) when $T_p = 933$ K] and (4) with two unknowns: $T_p$ (or $f_s$) and $m_{ox}$. All other parameters are explicitly expressed through $T_p$, $f_s$, and $m_{ox}$. In the Fuchs model, the algebraic equation, flux matching condition, Eq. (25), with the third unknown $T_\delta$ is added to the differential equations. For a fixed value of $T_\infty$, the differential equations (the algebraic-differential set in the case of the Fuchs model) were solved and maximum particle temperature $T_{p,max}$ was determined. Treating the maximum particle temperature as a function of the ambient temperature, we found the ignition temperature as the root of the equation

$$T_{p,max}(T_\infty) - T_{m,ox} = 0. \quad (31)$$

$T_{p,max}$ is a monotonous function of ambient temperature $T_\infty$, and Eq. (31) was solved by the bisection method.

TABLE I. Values of parameters used in calculations.

| Quantity | Value | References |
| --- | --- | --- |
| Density $\rho_{ox}$ | 3990 kg/m$^3$ | 39 |
| Density $\rho_{Al,l}$ | 2377 kg/m$^3$ | 39 |
| Density $\rho_{Al,s}$ | 2700 kg/m$^3$ | 39 |
| Emissivity $\epsilon$ | 0.3 | 40 |
| Melting point $T_{m,Al}$ | 933 K | 39 |
| Melting point $T_{m,ox}$ | 2327 K | 39 |
| Preexponent $k_0$ | 7.6 × 10$^{-6}$ kg/m s | 10 |
| Activation energy $E$ | 71 kJ/mol | 10 |

## IV. RESULTS

### A. Continuum model

Continuum model of ignition with the parabolic oxidation law has been considered theoretically before, and the details are well known.[16,17] The main results are briefly presented here.

For large particles igniting in a continuum regime—i.e., having diameters of tens of micrometers and larger—radiation can become an important factor. Nevertheless, it is instructive to consider the problem excluding the effect of radiation to elucidate contribution of reaction heat and conductive heat loss. Figure 1 presents the results of continuum model calculations: ignition temperature for micron-sized particles with two different initial oxide layer thicknesses: 3 nm and 10 nm. Both cases with radiation and without radiation ($\epsilon = 0$) are shown.

If radiation is neglected, as Fig. 1 shows, for particles larger than approximately 10 $\mu$m diameter, ignition temperature does not depend on the particle size and initial oxide thickness. For smaller particles, ignition temperature is a decreasing function of a particle diameter, and for a fixed particle size, ignition temperature is higher for thicker oxide. Also, the thicker the initial oxide layer is the greater the particle diameter is corresponding to the onset of the constant ignition temperature dependence.

The effect of radiation is manifested in the increase in heat loss and ignition temperature for larger particles and is negligible at smaller sizes. As a result, a minimum is observed in ignition temperature dependence on the particle diameter (Fig. 1).

### B. Free-molecular model

The free-molecular model was calculated for particles from 20 nm to 10 $\mu$m diameters. As will be shown below, significant part of this range may belong to the transition region. But even then, the free-molecular mechanism can be the important contributor to the total heat transfer. Thus, the analysis of this model is useful also for micro-sized particles.

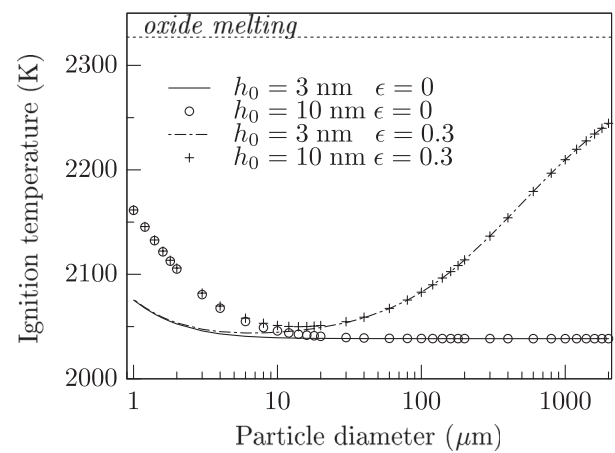

FIG. 1. Ignition temperature vs particle diameter for two initial oxide thicknesses $h_0$: 3 nm and 10 nm. The results with radiation ($\epsilon = 0.3$) and neglecting radiation ($\epsilon = 0$) are shown. Continuum model.

Figure 2 presents ignition temperature as a function of particle diameter at various values of the accommodation coefficient and initial oxide of thickness 3 nm. Ignition temperature, unlike predicted by the continuum model, increases with the particle size, except for small particles of several tens of nanometers. Two types of ignition were obtained. In the first type, particles are gradually heated to the oxide melting point. This ignition occurs for microsized particles and for nanoparticles when the accommodation coefficient is large, approximately $\alpha > 0.5$. The second type of ignition is typical for nanoparticles with sufficiently low values of accommodation coefficient. In this case, the critical ambient temperature can be defined: when the ambient temperature is lower than the critical value, a particle slowly oxidizes; at higher ambient temperatures, the particle ignites with the thermal runaway. Temperature behavior is similar to the critical ignition for the linear oxidation law, with some distinctions caused by the presence of protective oxide. (More details are given below.) The approximate boundary between the regions corresponding to these ignition modes is marked in Fig. 2. For smaller values of the accommodation coefficient, the critical ignition region extends to larger diameters and includes micrometer sizes.

Figures 3 and 4, where the maximum particle temperature is plotted against the ambient temperature $T_\infty$ for various particle diameters, illustrate the approach to the critical ignition with the change of particle size. In Fig. 3, the case of $\alpha = 1$ is shown. For large particles, maximum temperature dependence on ambient temperature is close to the linear function, while for smaller particles, the slopes of the curves increase with the ambient temperature. The smaller the particle is the steeper the slopes are.

In the case of $\alpha = 0.05$ (Fig. 4), the slope becomes virtually infinite for particles from about 1 $\mu m$ diameter and less. The transition from low temperature oxidation to ignition occurs within fractions of a kelvin of change of $T_\infty$. In a strict mathematical sense, no critical conditions exist for the cases shown in Fig. 4. Though some discontinuities in solutions on the vertical parts of the curves were obtained for small values of accommodation coefficients, $\alpha = 0.01$, they can be attributed to the precision limitations of numerical solution. Nevertheless, for all practical purposes, ignition of nanoparticles in Fig. 4 has critical behavior. The boundary that outlines the critical ignition region in Fig. 2 was calculated as follows. First, for fixed values of $\alpha$ and initial particle diameter, the ignition temperature $T_i$ is found. If the increase in this temperature by 1 K leads to the increase in maximum particle temperature by more than 1000 K, the case is considered as critical ignition.

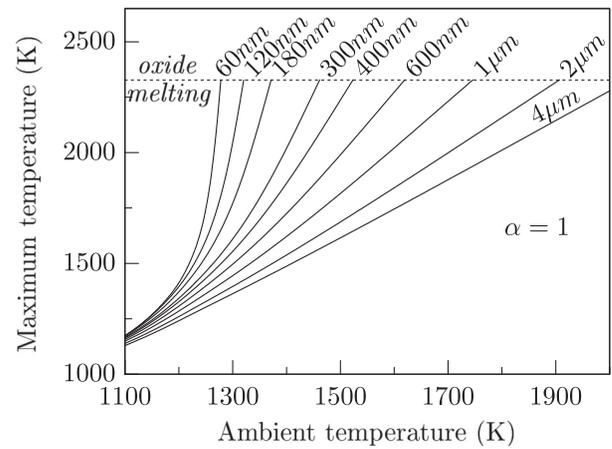

FIG. 3. Maximum particle temperature as a function of ambient temperature for various values of the particle diameter. $\alpha = 1$ and $h_0 = 3$ nm. Free-molecular model.

Figures 5 and 6 show two typical cases of particle temperature change with time and corresponding Semenov's diagrams. Cases of three different ambient temperatures are presented: below ignition temperature, equal to ignition temperature, and higher than ignition temperature. Time evolution patterns of particle temperature include heating (with the plateau at 933 K, aluminum melting) to some maximum temperature and consequent slow cooling to the environmental temperature. The gradual particle heating is shown for $\alpha = 1$ and a particle diameter of 600 nm in Fig. 5(a). The difference between the ambient temperatures for the curves is

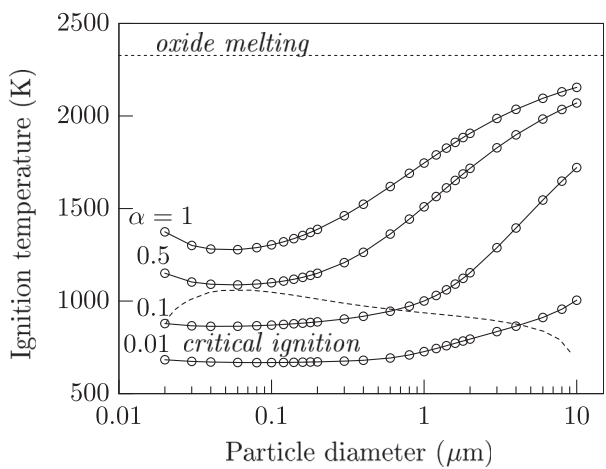

FIG. 2. Dependence of ignition temperature on the particle diameter for various values of the accommodation coefficient and the initial thickness $h_0 = 3$ nm. The dashed line delineates the region where ignition shows critical behavior. Free-molecular model.

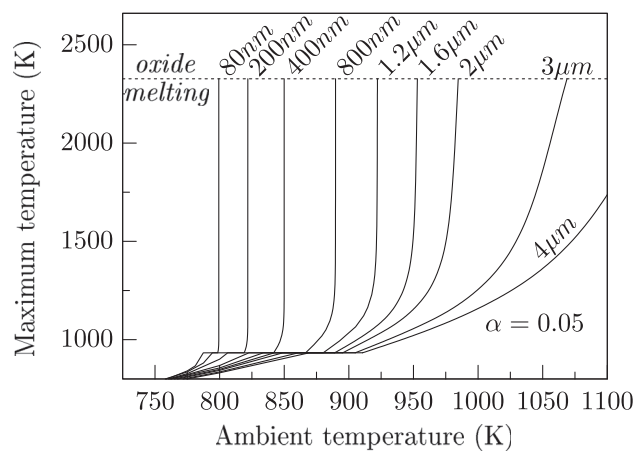

FIG. 4. Maximum particle temperature as a function of ambient temperature for various values of the particle diameter. $\alpha = 0.05$ and $h_0 = 3$ nm. Free-molecular model.

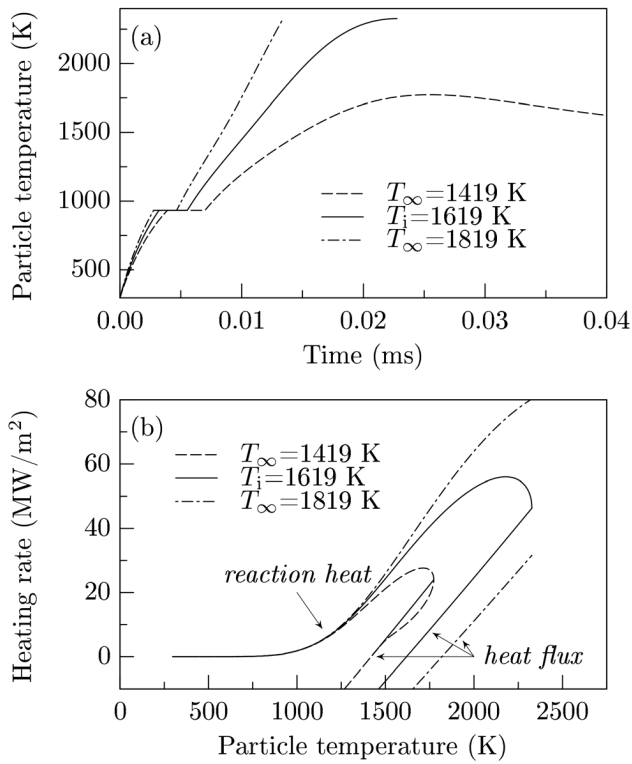

FIG. 5. Gradual heating to the melting point of oxide. (a) Particle temperature change and (b) Semenov's diagram for various ambient temperatures. Initial thickness $h_0 = 3$ nm, accommodation coefficient $\alpha = 1$, and particle diameter $d_p = 600$ nm. Free-molecular model.

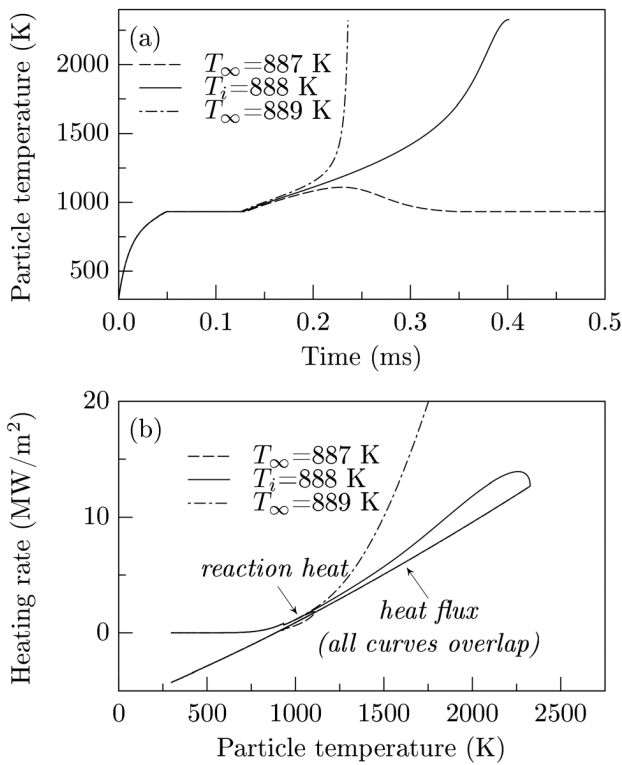

FIG. 6. Critical ignition. (a) Particle temperature change and (b) Semenov's diagram for various ambient temperatures. Initial thickness $h_0 = 3$ nm, accommodation coefficient $\alpha = 0.1$, and particle diameter $d_p = 200$ nm. Free-molecular model.

200 K. There is no thermal runaway even when $T_\infty$ is 200 K higher than ignition temperature.

The curves on Semenov's diagrams represent reaction heat release rate densities and total heat fluxes at the particle surface as functions of particle temperature. The heat release rate is governed in the course of particle heating by two major factors: exponential increase in the diffusion coefficient with temperature and diffusional resistance of the growing oxide. The competition between these factors defines the maximum temperature of a particle, unlike in the case of linear oxidation where maximum temperature is defined solely by the heat release—heat loss balance. Particle temperature is limited by the oxide growth. This is illustrated in Fig. 5(b). Higher ambient temperature leads to a higher heating rate and thinner oxide layer. Hence, a particle is heated to higher temperature before oxide becomes thick enough to cause the heat release drop [decreasing part of the heat release curve in Fig. 5(b)].

Figure 6 ($\alpha = 0.1$ and diameter 200 nm) illustrates the critical ignition case. The difference in ambient temperatures for the curves shown in Fig. 6(a) is only 1 K: 887 K, 888 K, and 889 K. At 887 K, the particle reaches maximum temperature only slightly higher than 1000 K. An increase by only 1 K in ambient temperature allows the particle to reach the oxide melting point. When ambient temperature increases again by 1 K, the thermal runaway occurs. Note that with the decrease of the accommodation coefficient, the ignition time scales increase accordingly: on the order of 0.01 s in Fig. 5(a) and 0.1 s in Fig. 6(a). In the case of $\alpha = 0.1$, the thermal runaway occurs on a time scale comparable with the ignition delay in the case of $\alpha = 1$.

The diagram in Fig. 6(b) is similar to the one in the Semenov thermal ignition model.[41] The cases shown differ only by 1 K ambient temperature, and all the heat flux curves overlap. As mentioned above, there is an important difference with the Semenov's ignition: in the current case, the thermal runaway is the result of domination of diffusion coefficient growth over oxide resistance. The heat release and heat loss curves approach each other very closely, but no conditions exist when the heat flux curve is a tangent to the reaction heat curve.

Qualitatively, the ignition temperature trends in Fig. 2 can be interpreted from the point of view of the heat release and heat loss dependence on the particle size at the moment of ignition. Using $r_{Al} = r_p - h$, we have for the heat release rate at the ignition moment

$$\frac{\dot{q}_{ri}}{r_p^2} \propto \frac{1}{h_i} - \frac{1}{r_p}, \quad (32)$$

where the subscript $i$ refers to the values at the ignition moment. Because both $\dot{q}_{fm}$ and radiation heat flow rates are proportional to $r_p^2$, the total heat loss flux does not depend on the particle size. For relatively large particles, the term $1/r_p$ in Eq. (32) can be neglected. The greater the particle size is, the greater the $h_i$ is and the lower the heat release rate density is. Hence, the ignition temperature increases with the particle size. For small particles when the oxide thickness becomes comparable to the particle size, the term $1/r_p$ in Eq. (32)

cannot be neglected. $h_i$ changes insignificantly and the reaction heat rate density at the ignition moment changes with the particle size mostly due to the $1/r_p$ term. The result is higher heat release rate density and lower ignition temperatures for larger particles. This is the effect of diffusion in spherical geometry expressed by the core radius term $r_{Al}$ in Eq. (4). The effect is observed in calculations for particle diameters smaller than approximately 30–50 nm when the initial oxide thickness is 3 nm and for diameters smaller than 100 nm when the initial oxide thickness is 10 nm.

Figure 7 shows ignition temperatures for two initial thicknesses: $h_0 = 3$ nm and $h_0 = 10$ nm. Except for large particle sizes, ignition temperatures are higher for larger values of $h_0$. The particle sphericity effect (decreasing temperature trend with the increasing particle size) is extended to larger particle sizes when the initial oxide layer is thicker. Particles with $h_0 = 10$ nm and diameters less than 30 nm are oxidized completely before temperature reaches the oxide melting point.

A word of explanation is necessary with regard to the radiation in the free-molecular regime. While for nanoparticles the emissivity depends on a particle size, all the calculations above were performed in the assumption of a constant particle emissivity. As calculations show, this assumption does not introduce significant error, at least for the cases considered here: the contribution of the radiation term to the total heat flux for nanoparticles, up to 1 $\mu$m diameter, is negligible. For example, the maximum ratio of radiation to conductive heat flow is on the order of 0.01 for $\alpha = 1$ at the maximum temperature, alumina melting point. For weak heat accommodation, contribution of radiation may become comparable to the heat conduction at 2300 K, but in this case, typically, the critical ignition occurs, and more relevant is comparison of heat flows at the maximum subcritical particle temperature. For example, for $\alpha = 0.01$, the ratio of radiation to conduction heat flux is about 0.05 (particle temperatures around 1000 K). It can be expected that the more precise account of emissivity of nanoparticles as proportional to the particle radius, $\epsilon \propto r_p$, would lead to even smaller contribution of radiation.

### C. Fuchs model

Ignition temperature resulting from the Fuchs model is presented as a function of particle diameter in Fig. 8 for initial thicknesses $h_0 = 3$ nm and $h_0 = 10$ nm and various values of accommodation coefficients. The results of free-molecular and continuum models are also shown in the figure. In the range of particle diameters, corresponding to the transition heat transfer regime, the Fuchs model yields lower ignition temperatures than those obtained by both free-molecular and continuum models. This range, as can be estimated visually from the figure, spans about two orders of magnitudes: e.g., from about 300 nm to 30 $\mu$m in the case of $\alpha = 1$. The decrease in the accommodation coefficient shifts the transition interval towards larger particle sizes. The initial thickness has no or minor effect on its boundaries.

For $\alpha = 0.01$, ignition temperatures obtained from the free-molecular model are close to the results of the Fuchs model, and for large particles igniting in a continuum regime, they are also close to the continuum model results. The cause of this is the increasing role of radiation for large particle sizes and small accommodation coefficients in a free-molecular model.

As was mentioned above, equations of the Fuchs model asymptotically approach the continuum ($Kn \to 0$) and free-molecular ($Kn \to \infty$) equations. The temperature at the Knudsen layer interface $T_\delta$, correspondingly, tends to $T_p$ and $T_\infty$. Therefore, the ratio $(T_p - T_\delta)/(T_\delta - T_\infty)$ can serve as a measure of contribution of free-molecular and continuum mechanisms to the heat transfer process. Substituting $r_\delta$ from Eq. (26) into the expression for continuum heat flow rate, Eq. (24), and equating the expression to the free-molecular flow rate Eq. (22), we obtain

$$\frac{T_p - T_\delta}{T_\delta - T_\infty} = \frac{Kn + Kn^2}{\kappa\alpha}, \qquad (33)$$

where the non-dimensional parameter $\kappa$ is

$$\kappa = \frac{pR_g^{1/2}\lambda}{2(2\pi T_\delta)^{1/2}k}\frac{\gamma^* + 1}{\gamma^* - 1}. \qquad (34)$$

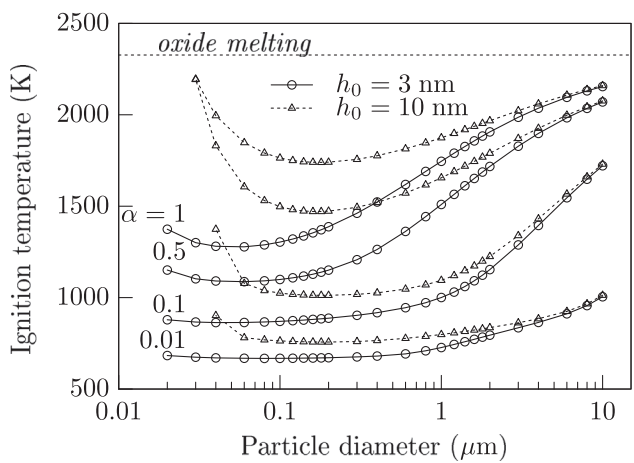
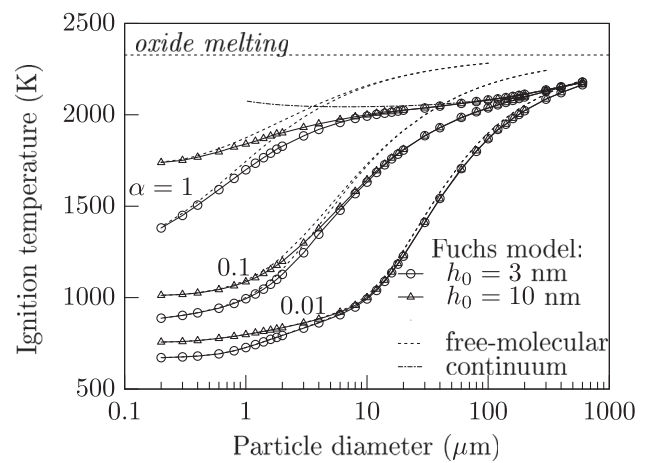

FIG. 7. Dependence of ignition temperature on the particle diameter for various values of accommodation coefficients and initial thicknesses $h_0 = 3$ nm and $h_0 = 10$ nm. Free-molecular model.

FIG. 8. Dependence of ignition temperature on the particle diameter for various values of accommodation coefficients and two initial thicknesses $h_0$: 3 nm and 10 nm. Comparison of all three models.

$(Kn + Kn^2)/(\alpha\kappa)$ is the ratio of free-molecular and continuum contributions to the transition heat transfer. If we expand Eq. (34) using the mean free path expression [Eq. (27)]

$$\kappa = \frac{k(T_\delta)}{k(T)} \frac{\gamma(T_\delta) - 1}{9\gamma(T_\delta) - 5} \frac{\gamma^* + 1}{\gamma^* - 1}, \quad (35)$$

we can see that $\kappa$ does not depend on pressure and depends on $T_\delta$ and $T$ only through the specific heat ratio $\gamma$ and heat conductivity $k$. For estimation purposes, we can assume constant $k$ and $\gamma$. Taking $\gamma = 1.3$ and $(\gamma^* + 1)/(\gamma^* - 1) = 7.67$, we obtain $\kappa = 0.34$.

Equation (33) can be used to determine the limits of different heat transfer regimes: $(Kn + Kn^2)/\alpha\kappa \gg 1$—free-molecular, $(Kn + Kn^2)/\alpha\kappa \ll 1$—continuum, $(Kn + Kn^2)/\alpha\kappa \approx 1$—transition heat transfer. Often, transition heat transfer regime limits are estimated by Knudsen numbers[36] having values between 0.01 and 100: $0.01 < Kn < 100$. By analogy, we can use as a criterion $0.01 < (Kn + Kn^2)/\alpha\kappa < 100$. These expressions can be written in terms of Knudsen number. Equating the right hand side of Eq. (33) to $\nu = 0.01$ and $\nu = 100$ and solving for $Kn$, we have for the limiting Knudsen numbers

$$Kn = \frac{(1 + 4\nu\alpha\kappa)^{1/2} - 1}{2}, \quad (36)$$

where $\nu = 0.01$ and $\nu = 100$ correspond to the continuum and free molecular regime limits, respectively. Using $\kappa = 0.34$ and estimating limiting values of $Kn$ from (36), we obtain the transition heat transfer limits as $3.4 \times 10^{-3} < Kn < 5.4$ for $\alpha = 1$ and $3.4 \times 10^{-5} < Kn < 0.27$ for $\alpha = 0.01$. These estimates show that using only the Knudsen number as a criterion for the heat transfer regime can be misleading.

Figures 9 and 10 illustrate how temperatures of the particle and the Knudsen layer interface change with time according to different models. Figure 9(a) shows heating of a 100 nm diameter particle when $\alpha = 1$. Heating occurs mostly according to the free-molecular heat transfer, with $T_\delta$ staying almost constant with the values close to $T_\infty$. Some minor difference between $T_\delta$ and $T_\infty$, several kelvins, occurs when the particle reaches the oxide melting point. The Knudsen number in this case has the values around 8. An example of transition heat transfer is presented in Fig. 9(b) for a particle of 4 $\mu$m diameter ($Kn \approx 0.2$–0.4), when both free-molecular and continuum temperatures differ significantly from those of the Fuchs model. The Knudsen layer interface temperature differs significantly from both $T_p$ and $T_\infty$.

The case of thermal runaway is presented in Fig. 10 for a relatively large particle, 800 nm diameter, and $\alpha = 0.05$ ($Kn \approx 0.65$). In Fig. 10(a), ignition at ambient temperature in the vicinity of the critical ignition temperature $T_i$ is shown. The particle is heated mostly in a free-molecular regime. $T_\delta$ stays close to $T_\infty$ and rises to the values several kelvins higher than $T_\infty$. Because the ambient temperature is close to the critical temperature $T_i$, this slight rise becomes important at times near the ignition moment and leads to earlier thermal runaway in the Fuchs model. This effect is less pronounced when the particle ignites at ambient temperatures much higher than critical temperature ($Kn \approx 1.2$), as shown in Fig. 10(b) for $T_\infty = 1500$ K. These examples illustrate that, despite the small values of Knudsen numbers and relatively large particle size, nanoparticle heating occurs mostly in a free-molecular regime.

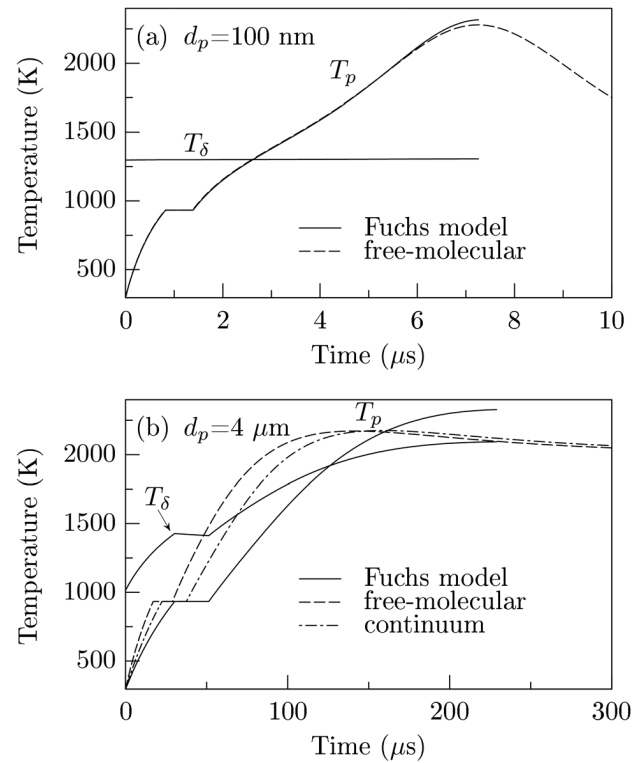

FIG. 9. Particle ($T_p$) and Knudsen layer interface ($T_\delta$) temperature change with time: (a). $\alpha = 1$, $d_p = 100$ nm, and $T_\infty = 1301$ K; (b). $\alpha = 1$, $d_p = 4$ $\mu$m, and $T_\infty = 1920$ K.

The comparison of the Fuchs model results with some experimental data from the literature is shown in Fig. 11 where ignition temperature vs. particle diameter is plotted. Experimental data are the results of compilation of many experiments performed under different conditions: environment, sample shape, heating rate, etc. This partly can explain the big data scattering. Theoretical ignition temperatures are shown for initial thickness $h_0 = 3$ nm and various values of accommodation coefficients. As Fig. 11 shows, the results of the Fuchs model are in qualitative agreement with the experimental ignition temperature trends.

## V. DISCUSSION

Historically, failure of continuum thermal ignition theory to explain the experimental dependence of ignition temperature on the particle size initiated the detailed study of reaction kinetics and other related processes inside particles, such as phase transformations or stresses developed in the oxide layer. The results of current simulations show that this dependence can be interpreted by transitional character of heat transfer: smaller particles ignite at lower ambient temperatures because of less efficient free-molecular heat removal from the particle. The ratio of free-molecular and continuum heat transfer contributions defines particle sizes, when this effect becomes noticeable. For low accommodation coefficients, these sizes can be relatively large, on the order of tens of micrometers.

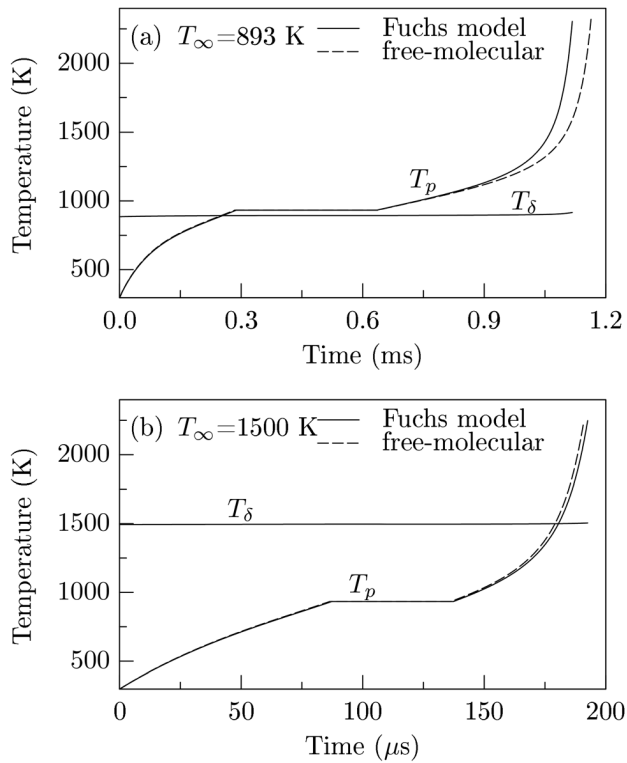

FIG. 10. Particle ($T_p$) and Knudsen layer interface ($T_\delta$) temperature change with time: (a). $\alpha = 0.05$, $d_p = 800$ nm, and $T_\infty = 893$ K; (b). $\alpha = 0.05$, $d_p = 800$ nm, and $T_\infty = 1500$ K.

Two possible types of ignition were found: ignition caused by gradual heating to the temperature at which the oxide loses protective properties and thermal runaway as a result of heat imbalance (critical ignition). The last type takes place in the case of small particles (roughly, nanoparticles, though the size range may include micrometers) and small accommodation coefficients ($\alpha < 0.1$). While oxide melting as an ignition criterion was used in both cases, in the case of critical ignition, this can be considered just a matter of convenience, partially, because no well-defined boundaries between two types of ignition were found. Ignition in this case is not necessarily related to the oxide melting.

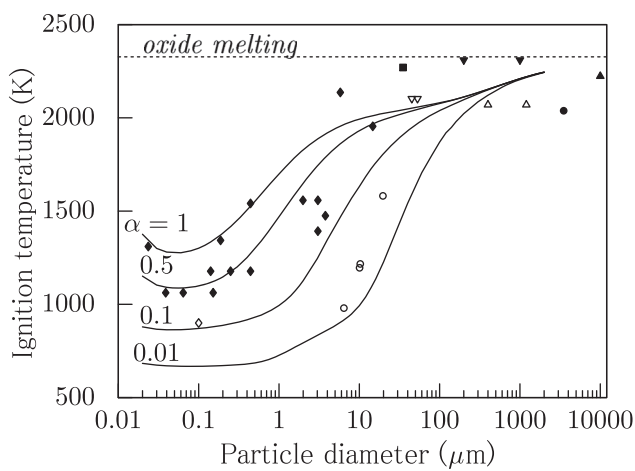

FIG. 11. Experimental and theoretical dependence of ignition temperature on the particle diameter. Theoretical data were obtained from the Fuchs model for various values of accommodation coefficients and the initial thickness of $h_0 = 3$ nm. Experimental data: Bulian et al.[5] ($\diamond$), Friedman and Maček[7] ($\blacksquare$), Derevyaga et al.[11] ($\bullet$), Ermakov et al.[13] ($\triangle$), Brossard et al.[14] ($\triangledown$), Yuasa[15] ($\blacktriangle$), Gurevich et al.[18] ($\circ$), Parr et al.[19] ($\blacklozenge$), Assovskiy et al.[42] ($\blacktriangledown$).

For particle sizes of several tens of nanometers, when the oxide thickness becomes comparable with the particle diameter, the ignition temperature decreases with the particle size increase. This is the effect of diffusion through the spherical oxide layer. There are some experimental evidences of similar dependence published in the literature, for ignition of aluminum[18] and carbon.[43]

Some experimental results that can be treated as ignition of a single nanoparticle were published in the literature. Parr et al.[19] reported ignition in a flame burner in the temperature range of 1000–1500 K that matches well the current calculations with $\alpha \approx 0.5$ (Fig. 11). Bazyn et al.[44] ignited nanoparticles by a shock wave at temperatures starting from 1200 K and elevated pressures of 4–32 atm. In experiments by Park et al.,[24] particles were subjected to the hot atmosphere over 1 s interval and even at temperatures as high as 1100 °C were oxidized only partially. Levitas et al.[27] suggested that at a high heating rate in shock waves, the melt dispersion mechanism is responsible for particle ignition, while at a low heating rate, slow diffusional oxidation occurs. The current results indicate that different heat transport conditions in a shock wave and hot atmosphere experiments and possibly, energy accommodation, can also be responsible.

The calculations showed that during melting, the ambient temperature is lower than the melting point. For example, for $\alpha = 0.01$, and 200 nm diameter, according to the calculations, the particle melts and ignites at the ambient temperature 683 K. Similar ignition or fast oxidation temperatures lower than the melting point were observed in some thermogravimetric measurements.[6] Park et al.[24] note that the results in these experiments can be affected by the heat and mass transfer, whereas single particle oxidation experiments are clear of such influences. The current simulations show that heat transfer is also an important factor in ignition of a single particle and may lead to much higher particle temperatures compared to the ambient temperature, up to 1000 K and higher. Neglecting this effect may lead to erroneous interpretations of experiments.

In the case of oxidation with polymorphic transformations in alumina by Trunov et al.,[22] the oxidation is diffusion limited, although with different parameters for different aluminum polymorphs, which is expected to lead to qualitatively similar results as in this study. The exposure of the metal surface to the oxidizer due to density differences of polymorphs plays the important role in the model. This leads, according to the model, to the high rate aluminum surface oxidation limited by the oxygen diffusion through the gas. The continuum heat and mass transfer was considered. If free molecular effects were taken into account, this effect could be expected to be less pronounced because of slower diffusion, especially taking into account reduction of sticking coefficients of oxygen on metal surfaces with increasing temperature.[45] A similar reasoning can be applied to the case of oxide film cracking caused by thermo-mechanical stresses. The use of these complex models could be justified if more precise kinetics was sought, but currently it is not feasible because accommodation coefficients are not known.

To separate the effects of oxidation kinetics and heat transfer, knowledge of the thermal accommodation coefficient is necessary. This may be important not only for nanoparticles but also for micron-sized particles, igniting or burning in a transition regime. Although a vast literature exists on energy accommodation in different gas-solid systems,[46] the conditions in these systems are very far from those of interest in combustion. To the best of our knowledge, there are no studies available in the literature on thermal accommodation of oxygen and nitrogen on metal oxide surfaces at elevated temperatures. Probably, the only exception is the work of Altman et al.[47] who studied the silica nanoparticles at conditions close to those existing in combustion (temperatures around 2000 K) and found that the accommodation coefficient does not exceed the value of 0.005.

## VI. CONCLUSIONS

Thermal ignition of an aluminum particle in the air was considered using continuum, free-molecular, and Fuchs heat transfer models. A single parabolic oxidation kinetics was assumed in the entire particle size range, from nano- to millimeter diameters. Ignition temperature dependence on the particle size obtained by the Fuchs model is in good qualitative agreement with the experimental trends. The major results can be summarized as follows:

(i) The limits of the transition heat transfer regime are defined by the ratio of free-molecular and continuum heat transfer, which may significantly differ from those estimated with the Knudsen number.
(ii) Micro-sized particles—up to tens or hundreds of micrometer diameters—ignite in a transition heat transfer regime. This may be the cause of experimentally observed decline of ignition temperature for smaller particles in the micrometer size range.
(iii) For small accommodation coefficients, nanoparticle ignition is governed mostly by the free-molecular heat transfer mechanism.
(iv) Depending on the thermal accommodation coefficient and size, nanoparticles may ignite either by gradual heating to a temperature at which oxidation kinetic changes or in a critical ignition mode with the thermal runaway.
(v) Regardless of an immediate cause of ignition, heat transfer is an important factor: particle temperature may greatly differ from environmental temperature. Neglecting this difference may lead to erroneous interpretation of experiments.
(vi) Thermal accommodation coefficient is a critical parameter for understanding the heat transfer effects in ignition of micro- and nanoparticles.

## APPENDIX: COMPARISON OF REACTION HEAT AND HEAT TRANSFER RATES

The reaction heat contribution can be evaluated by the following dimensionless parameters: the reaction heat to the continuum heat flow ratio in the continuum regime

$$\kappa_{r-c} = \frac{r_p}{k\Delta T} \frac{\dot{q}_{r,max}}{4\pi r_p^2} \quad (A1)$$

and the reaction heat to the free-molecular heat flow ratio in the transition regime

$$\kappa_{r-fm} = \sqrt{\frac{8\pi T_\infty}{R_g}} \frac{\dot{q}_{r,max}}{4\pi r_p^2 p \alpha \frac{\gamma^* + 1}{\gamma^* - 1} \Delta T}, \quad (A2)$$

where $\dot{q}_{r,max}$ is the maximum heat release rate, and $\Delta T$ is the maximum spatial temperature variation, i.e., the difference between particle and gas temperatures. Let us estimate these parameters for the ignition cases and initial thickness $h_0 = 3$ nm, for which the heat release rates are the strongest. In both continuum and transition regimes, $\Delta T$ has the order of $10^3$ K. According to the calculations performed, at ignition temperatures, the values of $\dot{q}_{r,max}/4\pi r_p^2$ are of the order of $10^5$ W/m$^2$ for large particles igniting in the continuum regime and increase to $10^7$ W/m$^2$ for particles igniting in the transition regime. Evaluation of Eqs. (A1) and (A2) shows that $\kappa_{r-c} \sim \mathcal{O}(1)$ and $\kappa_{r-fm} \sim \mathcal{O}(1)$ or less. The reaction heat is of the same order of magnitude or less than conductive heat flow.